\documentclass[11pt]{article}

\usepackage[T1]{fontenc}
\usepackage{lmodern}
\usepackage[utf8]{inputenc}
\usepackage[a4paper,margin=1in]{geometry}
\usepackage{amsmath,amssymb,amsthm}
\usepackage{booktabs,tabularx,array}
\usepackage{graphicx}
\usepackage[table]{xcolor}
\usepackage{enumitem}
\usepackage[round,authoryear]{natbib}
\usepackage{caption}
\usepackage{microtype}
\usepackage[section]{placeins}
\usepackage{tikz}
\usepackage{pgfplots}
\pgfplotsset{compat=1.17}
\usetikzlibrary{positioning,arrows.meta}
\usepackage[hidelinks]{hyperref}

\setlist{itemsep=2pt,topsep=4pt}

\newcolumntype{L}{>{\raggedright\arraybackslash}X}
\newcolumntype{P}[1]{>{\raggedright\arraybackslash}p{#1}}
\theoremstyle{plain}
\newtheorem{proposition}{Proposition}
\newtheorem{corollary}{Corollary}
\theoremstyle{definition}
\newtheorem{definition}{Definition}
\newtheorem{assumption}{Assumption}

\newcommand{\cp}{C\&P Test}

\title{Measuring the Human Contribution When AI Is Allowed:\\
Psychometric Foundations for Work-Sample Tests of AI Use}

\author{CH Peh\thanks{Corresponding author: chp@straitsai.com}, Jennifer Dodgson, Roger Sim\~{o}es, Shawn Chia, Elijah Ng}

\date{November 2026}

\begin{document}
\maketitle

\begin{abstract}
Employers increasingly need to know how well a person does their work with an AI assistant. A test of this has to let candidates use a live AI, and that creates a problem psychometrics has met before only in weaker forms: the tool may complete the task by itself. We review how measurement has handled tools in the testing room, from open-book examinations and calculators to computer-based simulations and scripted collaborators, and survey attempts to assess AI literacy and AI use, which have relied mostly on self-report and knowledge items. We then develop a measurement framework for questions on which AI is allowed. If a candidate who cannot do the human part of a task falls back on the cheapest strategy available (sending the AI's own output), the item response function takes the three-parameter logistic form, with a lower asymptote equal to that strategy's success rate. We call this the item's \emph{AI floor}. We show that the floor multiplies the item's information about the person by $1-c/P(\theta)$; that it can and should be estimated before an item is used, by simulation with the test's own pinned model, with exact binomial bounds that fix the number of runs required; that it must be estimated against the strongest cheap strategy, not the plainest; that it must be re-estimated after any change of model or of the item's surface; and that the AI's own randomness enters generalisability theory as a new source of error. For scoring, we ground authorship-neutral marking of the delivered work in accountability and in the distinction between effects \emph{with} and effects \emph{of} a technology, measure the person's own understanding with the AI switched off, model checking tasks with signal detection theory, grade each item by ordered, conjunctive bands of learning objectives, and treat serious failures as ceilings rather than deductions. We illustrate the framework with the design of the SIAA Standard; with about 700 simulated runs, which show that, with a current model, many familiar ways of making a task hard are neutralised by the AI alone; and with grading tests in which one plain copier in 155 reached the passing standard.
\end{abstract}

\section{Introduction}

AI assistants are changing what a job requires. A person who uses an assistant well can do more, and a person who leans on it for work beyond its abilities can do worse than without it \citep{dellacqua2026jagged}; across many experiments, people working with AI beat people working alone on average, but often fall short of the better of the two working alone \citep{vaccaro2024combinations}. Employers therefore want evidence of a skill that existing credentials do not show: how well this person does real work with AI, and whether their AI-assisted work can be trusted.

Measuring that skill directly means letting the candidate use an AI during the test. Psychometrics has admitted tools to the testing room before: books, calculators, computers, simulated colleagues. Each time, the construct changed from what a person can do unaided to what they can do with the tool, and items had to be redesigned so that the tool did not do the work alone. A large language model differs from these earlier tools in one decisive way. A calculator computes but does not decide what to compute; a book holds facts but does not apply them. An AI assistant, given the task, can often produce a complete and correct answer. A candidate who pastes the task into the assistant and sends back its reply may then score as well as a skilled user, and the score measures the AI \citep{messick1995validity,haladyna2004cirv}.

This paper sets out a psychometric foundation for tests in which AI is allowed. Section~\ref{sec:primer} summarises the measurement theory the argument needs. Section~\ref{sec:history} reviews how assessment has handled tools, simulations and collaborators. Section~\ref{sec:attempts} reviews attempts to assess AI literacy and AI use. Section~\ref{sec:theory} sets out the theoretical commitments: what the construct is, and why it must be observed in the joint work of person and AI. Section~\ref{sec:model} develops a measurement model for AI-allowed items, with the AI floor at its centre. Section~\ref{sec:scoring} develops the foundations for scoring. Section~\ref{sec:siaa} describes how the SIAA Standard puts the framework into practice, and Section~\ref{sec:evidence} reports simulation evidence. Section~\ref{sec:discussion} states the validity argument the framework must still support.

In keeping with practice for secure tests \citep{aera2014standards,wollack2013security}, operational questions, Answer Keys, the lists of learning objectives and serious failures, and scoring constants are not published here; examples are schematic, and constants appear as symbols.

\section{A short primer on the measurement theory used}
\label{sec:primer}

\subsection{Constructs and validity}

Early validity was a correlation: a test was valid for whatever it predicted. Content validity added the question of whether a test samples its domain, and construct validity, introduced for attributes with no single criterion, asked whether a pattern of evidence supports the interpretation of scores as measures of a theoretical attribute \citep{cronbach1955construct}. \citet{messick1989validity} unified these into one concept, construct validity, and \citet{messick1995validity} later set out six aspects of it: content, substantive (the processes candidates actually use), structural, generalisability, external and consequential.

A \emph{construct} is the attribute a test is meant to measure. Validity is not a property of a test but of the interpretation and use of its scores: the degree to which evidence and theory support them \citep{messick1989validity,aera2014standards}. \citet{messick1995validity} names two major threats to construct validity. \emph{Construct under-representation} means the test leaves out part of the construct. \emph{Construct-irrelevant variance} means scores depend on something other than the construct; when that something makes items easier for some candidates without the skill, \citet{messick1995validity} calls it construct-irrelevant easiness. Test-wiseness and cheating are classic sources \citep{haladyna2004cirv}. Copying an AI's answer on a test of AI use is a new and unusually powerful source of construct-irrelevant easiness.

Modern validity theory frames validation as an argument \citep{kane2013validating}: a chain of inferences from an observed performance to a score (scoring), from the score on these tasks to performance on the universe of similar tasks (generalisation), from that to performance in the real domain (extrapolation), and from there to a decision. Each inference rests on assumptions that need evidence. We use this structure in Section~\ref{sec:discussion}.

\subsection{Reliability: classical test theory and generalisability theory}

Classical test theory writes an observed score as a true score plus error, $X = T + E$, with reliability $\rho = \sigma^2_T / \sigma^2_X$ \citep{lord1968statistical}. Lengthening a test by a factor $k$ with parallel items raises reliability according to the Spearman--Brown formula,
\begin{equation}
\rho_k = \frac{k\rho}{1+(k-1)\rho},
\label{eq:sb}
\end{equation}
which is why precision lost per item must be bought back with more items (Section~\ref{sec:model}). Generalisability theory \citep{cronbach1972dependability,brennan2001generalizability} replaces the single error term with several, one for each source of variation (facet) in the measurement design. For persons $p$ crossed with items $i$ and raters $r$, the observed score decomposes as
\begin{equation}
X_{pir} = \mu + \nu_p + \nu_i + \nu_r + \nu_{pi} + \nu_{pr} + \nu_{ir} + \nu_{pir,e},
\end{equation}
and the generalisability coefficient for comparing persons with $n_i$ items and $n_r$ raters is
\begin{equation}
E\rho^2 = \frac{\sigma^2_p}{\sigma^2_p + \sigma^2_{pi}/n_i + \sigma^2_{pr}/n_r + \sigma^2_{pir,e}/(n_i n_r)}.
\label{eq:gtheory}
\end{equation}
The most important empirical result for performance assessment came from this framework: the person-by-task interaction $\sigma^2_{pi}$ is large, so a person's score depends heavily on which tasks they happen to get, and many tasks are needed for a dependable score \citep{shavelson1993sampling}.

\subsection{Item response theory}

Item response theory models the probability that person $p$ with latent ability $\theta_p$ succeeds on item $i$ \citep{rasch1960,lord1968statistical,hambleton1991fundamentals}. The two-parameter logistic (2PL) model is
\begin{equation}
F_i(\theta) = \frac{1}{1+\exp[-a_i(\theta-b_i)]},
\end{equation}
with difficulty $b_i$ and discrimination $a_i$. The three-parameter logistic (3PL) model adds a lower asymptote $c_i$ \citep{birnbaum1968some}:
\begin{equation}
P_i(\theta) = c_i + (1-c_i)\,F_i(\theta).
\label{eq:3pl}
\end{equation}
The 3PL was introduced for multiple-choice items, where a person who does not know the answer can still guess it. It can be read through a ``knowledge or random guessing'' account: with probability $F_i(\theta)$ the person knows the answer; otherwise they guess, and succeed with probability $c_i$ \citep{vanderven1992krg,maccann2004krg}. In practice $c_i$ is often below the chance rate, so it is better called a pseudo-chance level than a guessing parameter \citep{hambleton1991fundamentals}. Section~\ref{sec:model} shows that the same structure arises, for a different reason, whenever AI is allowed.

Both models assume local independence: given $\theta$, responses to different items are independent. Items that share a stimulus, such as several questions about one case, violate it and are better treated as a single unit, a testlet \citep{wainer1987testlets}.

The precision an item contributes at ability $\theta$ is its Fisher information,
\begin{equation}
I_i(\theta) = \frac{[P_i'(\theta)]^2}{P_i(\theta)\,[1-P_i(\theta)]},
\label{eq:info}
\end{equation}
and test information is the sum over items. The standard error of an ability estimate is approximately $1/\sqrt{\sum_i I_i(\theta)}$. Computerised adaptive testing chooses each next item to maximise information at the current ability estimate, which concentrates use on the most informative items; exposure control limits how often any item is shown, for example by administering a selected item only with a probability tuned so that its exposure rate stays below a ceiling \citep{sympson1985exposure}. Exposure matters because an item that circulates among candidates no longer measures what it did.

\subsection{Evidence-centred design and item models}

Evidence-centred design describes an assessment through linked models, three of them central: a \emph{student model} (the claims to be made about a person), an \emph{evidence model} (what observations support those claims and how they are scored) and a \emph{task model} (situations that elicit the observations) \citep{mislevy2003structure}. Construct representation, the question of which cognitive processes an item actually calls on, is a separate validity question from how scores relate to other variables \citep{embretson1983construct}.

Automatic item generation builds many items from one \emph{item model} \citep{gierl2013items}. Features that change an item's difficulty are \emph{radicals}; features that should not are \emph{incidentals} \citep{irvine2002item,gierl2012aig}. Items that share radicals and differ only in incidentals are \emph{isomorphs}, intended to be psychometrically exchangeable; \citet{embretson1998cognitive} showed that an item's difficulty can be predicted from its design features.

\subsection{Fairness}

An item functions differentially if candidates of equal ability but from different groups have different chances of success. The Mantel--Haenszel procedure \citep{holland1988dif} matches a reference group and a focal group on total score, forms a $2\times2$ table of group by success in each score stratum $k$ (with cells $A_k, B_k$ for the reference group's successes and failures, $C_k, D_k$ for the focal group's, and $N_k$ candidates in all), and pools the odds ratios:
\begin{equation}
\hat\alpha_{\mathrm{MH}} = \frac{\sum_k A_k D_k / N_k}{\sum_k B_k C_k / N_k}, \qquad \mathrm{MH\ D\text{-}DIF} = -2.35\,\ln \hat\alpha_{\mathrm{MH}} .
\end{equation}
For a test of AI use, the groups of interest include language background, field of work and prior access to AI tools.

\subsection{Standard setting}

A cut score turns a scale into decisions. In the Angoff method and its variants, judges estimate, for each item, the probability $p_{ij}$ that a minimally competent candidate would succeed; the cut score is the sum over items of the judges' mean estimates \citep{angoff1971scales,cizek2007standard}:
\begin{equation}
C = \sum_i \bar p_i, \qquad \bar p_i = \frac{1}{J}\sum_{j=1}^{J} p_{ij}.
\end{equation}
The reference point is a hypothetical borderline person, not the best or the average. Section~\ref{sec:scoring} uses the same device to define a serious failure: a competent person working without AI.

\subsection{Test security}

Professional standards require that test content be protected, because exposure changes what scores mean \citep{aera2014standards,wollack2013security}. Secure programmes publish their frameworks, specifications and validity evidence but not operational items or Answer Keys. We follow that practice here.

\subsection{Raters}

Constructed responses need raters, and raters differ in severity. The many-facet Rasch model \citep{linacre1989mfrm} places persons, items, raters and score categories on one scale:
\begin{equation}
\log\frac{P_{nijk}}{P_{nij(k-1)}} = \theta_n - \delta_i - \lambda_j - \tau_k,
\label{eq:mfrm}
\end{equation}
where $P_{nijk}$ is the probability that rater $j$ gives person $n$ category $k$ on item $i$, $\delta_i$ is item difficulty, $\lambda_j$ rater severity and $\tau_k$ the threshold of category $k$. Rater severity can drift over time and needs monitoring \citep{hoskens2001drift}. Frameworks for evaluating automated scoring treat an automated rater as one more rater whose agreement, severity and fairness must be shown \citep{williamson2012framework}.

\section{Tools, simulations and collaborators in the testing room}
\label{sec:history}

\subsection{Open books and calculators}

The open-book examination is the oldest tool-allowed test. Research comparing open- and closed-book formats gives mixed results: a systematic review found that scores tended to favour closed-book exams, especially when students prepared more for them, and that students may prepare more for closed-book exams \citep{durning2016openbook}, while experiments found open-book scores higher at first but equal on a delayed test \citep{agarwal2008openbook}. The design lesson was that an open-book question must ask for something the book does not contain: application, not lookup.

Calculators forced a more explicit redesign. Research through the 1980s found that calculator use alongside traditional instruction did not harm students' paper-and-pencil skills, except at grade 4 \citep{hembree1986calculators}, and in 1994 the SAT began to permit calculators \citep{lawrence2003sat}. A study of calculator use on a test of mathematical reasoning found effects that varied from item to item \citep{bridgeman1995calculator}. Programmes responded by classifying items by their relation to the tool: the US National Assessment of Educational Progress classes mathematics items as calculator-inactive, calculator-neutral or calculator-active \citep{naep_calculator}. The principle carries directly to AI: once a tool is allowed, each item must be examined for what the tool can do on it alone.

\subsection{Work samples and performance assessment}

Work-sample tests ask candidates to do a sample of the job. In personnel selection they are among the more valid predictors of job performance, although estimates vary between meta-analyses \citep{schmidt1998validity,roth2005worksample}. Performance assessment in education gained ground around 1990, with explicit criteria for its validation \citep{linn1991complex,messick1994interplay,kane1999validating}. Its central difficulty proved to be generalisation: performance on one task predicts performance on another only weakly \citep{shavelson1993sampling}.

\subsection{Standardised counterparts}

When a performance involves another party, the other party must be standardised. The objective structured clinical examination (OSCE) moves candidates through a series of stations, each a set task, some marked by an examiner against a checklist \citep{harden1975osce}, and standardised patients, people trained to portray a patient in a standardised way, present the same history to every candidate \citep{barrows1993sp}. Computer-based case simulations in medical licensure present the patient in software rather than through an actor, and score the candidate's management actions automatically \citep{clauser1997automated}. Critical actions, checklist items essential to patient safety, were proposed as a separate check on OSCE results, since in one school's data many candidates with high overall marks missed them \citep{payne2008critical}; in licensure examinations, some critical errors fail the candidate whatever the other marks \citep{buckendahl2012dce}.

\subsection{Judgement in simulated situations}

Situational judgement tests present realistic work dilemmas and ask what the candidate would do; they predict job performance \citep{mcdaniel2001sjt}. They show that judgement about work can be tested, but they test it at one remove: the candidate chooses a described action rather than doing the work.

\subsection{Computers as the setting}

Performance tests of digital skill moved the candidate into simulated software. ETS's iSkills assessment asked students to find, evaluate and communicate information in simulated applications \citep{katz2007iskills}, and the International Computer and Information Literacy Study measured students' computer and information literacy through tasks in a simulated environment \citep{fraillon2014icils}. The OECD Survey of Adult Skills assessed problem solving in technology-rich environments: adults used simulated email, spreadsheet and web tools to complete tasks \citep{oecd2012piaac}. Here the tool was part of the construct. Logged actions became evidence, and process data from such tasks proved informative but in need of their own validation \citep{greiff2015logfile,goldhammer2021byproduct}.

\subsection{Collaborators}

The Assessment and Teaching of 21st Century Skills project \citep{griffin2012atc21s} defined collaborative problem solving as a set of social and cognitive skills \citep{hesse2015cps} and assessed it with online tasks taken by pairs of students \citep{griffin2015methods}. PISA 2015 used scripted computer agents instead of human partners, so that the partners behaved the same way for every student taking a given task \citep{oecd2017pisa2015}; a main stated reason was that consistency and control are essential to any assessment \citep{graesser2017cps}. A pilot study found no significant difference in problem-solving scores between agent and human partners, but higher collaboration scores with agents \citep{rosen2015h2a}, and a comparison using the PISA tasks found that agent partners gave scores comparable to human ones \citep{herborn2020agents}. A recent proposal uses one language model to play all of a candidate's teammates and steer the conversation to elicit the target skills \citep{globerson2026durable}.

\subsection{What changes with AI}

Each of these tools changed the construct, and each was handled by redesigning items so that the tool could not do the work alone, or by scripting the counterpart. A language model breaks both responses. It can often do the whole task alone, so the redesign must be far more thorough; and scripting it would remove the very behaviour, including its errors, that a person working with AI must handle. Table~\ref{tab:tools} summarises the progression.

\begin{table}[!htbp]
\centering\small
\caption{Tools in the testing room, and what each required of item design.}
\label{tab:tools}
\begin{tabularx}{\textwidth}{@{}P{2.6cm}LL@{}}
\toprule
Tool & What it can do alone & What items had to do \\
\midrule
Book & Supply facts & Ask for application, not lookup \\
Calculator & Compute & Classify items by their relation to the calculator; make it neutral or required \\
Simulated software & Present a realistic setting & Score logged actions; validate process indicators \\
Scripted collaborator & Behave the same for everyone & Script the partner to standardise conditions \\
AI assistant & Often, the whole task & Show before use that the AI alone does not pass; keep the AI unscripted but pinned \\
\bottomrule
\end{tabularx}
\end{table}

\section{Past attempts to assess AI literacy and AI use}
\label{sec:attempts}

\subsection{Frameworks}

Definitions of AI literacy came first. \citet{long2020literacy} set out competencies a non-expert needs to understand and use AI; \citet{ng2021conceptualizing} grouped the literature into knowing and understanding AI, using and applying it, evaluating and creating with it, and AI ethics. International frameworks followed: UNESCO's competency framework for students \citep{unesco2024students}, the OECD and European Commission's AI literacy framework \citep{oecd2025ailit}, and the planned PISA 2029 assessment of media and AI literacy \citep{oecd2029mail}. A framework developed by two academics for higher education, and taught in open courses produced with an AI company, describes AI fluency in four parts: delegation, description, discernment and diligence \citep{dakan2025fluency}. These frameworks describe what people should be able to do; they do not by themselves measure it.

European digital-competence frameworks have also been extended with examples of interacting with AI systems \citep{vuorikari2022digcomp}.

\subsection{Self-report scales}

Most instruments are self-report scales, such as the Meta AI Literacy Scale \citep{carolus2023mails} and the AI literacy scale of \citet{wang2023ails}. A systematic review found 16 AI literacy scales, only 3 of them performance-based \citep{lintner2024review}. Self-report measures what people believe about their skill, which is a different construct from the skill.

\subsection{Knowledge tests}

Knowledge tests ask what AI is and how it behaves. \citet{hornberger2023literacy} developed and validated such a test for university students, and GLAT assesses generative-AI literacy with multiple-choice items \citep{jin2025glat}. Knowing that models hallucinate is useful, but knowing it is not the same as catching a hallucination in one's own work under time pressure.

\subsection{Performance with AI}

Observational studies came first. Non-experts asked to design prompts relied on trial and error, generalised from single examples, and struggled to recover when outputs went wrong \citep{zamfirescu2023johnny}; in a study of university students, the quality of their prompts predicted the quality of what the model produced \citep{knoth2024prompt}. A small number of studies go further and measure people working with AI. \citet{riedl2025synergy} use item response theory to separate a person's ability alone from their ability with AI, and find the second to be a distinct trait. CollabSkill \citep{shao2026collabskill} observes workers using AI agents on tasks from their own occupations. ChatBench \citep{chang2025chatbench} shows that an AI's accuracy alone does not predict the accuracy of a person working with it. These are research instruments built on tasks that are not screened against the AI working alone; on many of them the AI alone does well. Commercial coding assessments now allow an AI assistant during the test, but we found no published validation of them. A recent review of measures for competent generative-AI use found, among the studies it reviewed, no validated individual-level instrument that tests the full set of skills for controlling AI agents, including setting their scope and permissions, recovering from errors and checking their work independently \citep{veri2026beyond}.

\subsection{AI as a threat to existing tests}

A parallel literature treats AI as a threat. Studies have run ChatGPT on existing examinations to find vulnerable assessments \citep{nikolic2023chatgpt,borges2024engineering,hinchcliffe2026vulnerability} and predicted which items are vulnerable \citep{ram2023vulnerability}. Responses include linked problems \citep{ding2025interconnected}, ``assessment twins'' in which a second task checks the first \citep{roe2025twins}, and the argument that AI use is a question of validity rather than cheating \citep{dawson2024validity}. AI benchmarks use a related mechanism in reverse: adversarial filtering swaps out wrong answer options that a model can easily spot until the questions are hard for models \citep{zellers2018swag}, and Humanity's Last Exam admitted questions only if leading models failed them \citep{hle2026nature}.

\subsection{The gap}

Table~\ref{tab:gap} sets the approaches side by side. None combines the three features a test of AI use needs: a live AI during the test, tasks that are real work, and evidence, before the task is used, that the AI alone cannot pass it.

\begin{table}[!htbp]
\centering\small
\caption{Approaches to assessing AI literacy and AI use.}
\label{tab:gap}
\begin{tabularx}{\textwidth}{@{}LP{1.8cm}P{1.8cm}P{3.0cm}@{}}
\toprule
Approach & Live AI in the test & Real work & Items screened against the AI alone \\
\midrule
Frameworks & No & --- & --- \\
Self-report scales & No & No & --- \\
Knowledge tests & No & No & --- \\
Research studies of human--AI performance & Yes & Sometimes & No \\
Vulnerability audits of existing exams & AI as threat & Sometimes & Yes, to keep AI out \\
AI benchmarks with adversarial filtering & AI is the subject & No & Yes, to measure the AI \\
This paper & Yes & Yes & Yes, to measure the person \\
\bottomrule
\end{tabularx}
\end{table}

\section{Theoretical commitments}
\label{sec:theory}

\subsection{Partners in cognition: effects with and effects of}

\citet{salomon1991partners} distinguished two kinds of effect of an intelligent technology on a person. \emph{Effects with} the technology are changes in performance while working in partnership with it: the joint system of person and tool does more than the person alone. \emph{Effects of} the technology are the cognitive residue the partnership leaves: what the person can do afterwards without it. The work on distributed cognition \citep{hutchins1995cognition} and the extended mind \citep{clark1998extended} goes further and treats people together with their tools as the unit of cognition.

We adopt this distinction as the basis of the measurement design. An employer receives the product of the joint system; that is the performance that matters, and it should be observed with the AI present. But a product that the person cannot explain is a product they cannot stand behind, correct or defend; whether they understand what they sent is a residue observed with the AI absent. A test of AI use therefore needs both kinds of observation, kept apart.

\subsection{Reliance, oversight and decision rights}

Research on human--automation interaction describes four ways people relate to automation: use, misuse (over-reliance), disuse (under-reliance) and abuse (automation imposed without regard to the people affected) \citep{parasuraman1997humans}. Good reliance is calibrated: trust matches the automation's actual reliability in the situation at hand \citep{lee2004trust}. Experiments on AI advice measure reliance from both sides, accepting advice when it is right and overriding it when it is wrong \citep{schemmer2023appropriate}, or focus on over-reliance, accepting advice when it is wrong \citep{bucinca2021trust}; and in one study, explanations from the AI raised acceptance of its recommendations whether they were right or wrong, and did not improve team accuracy beyond simply showing the AI's answer and confidence \citep{bansal2021whole}. Two consequences follow for a test. Catching AI errors must be measured together with leaving correct output alone; and a person must be able to tell which decisions are theirs to make and which belong to someone else, however confidently the AI offers to make them.

\subsection{The construct}

We define the construct as \emph{the person's contribution to work done with an AI assistant}: the difference their judgement, actions and knowledge of how to work with AI make to the quality and trustworthiness of the joint product, together with their ownership of that product. Knowledge of the job itself is not part of the construct: every fact the work needs is given in the item's materials, and the work is judged right or wrong against them. The construct is observed in two ways: in the joint product (effects with), and in the person's understanding of it without the AI (effects of, in a local sense). What the AI does without the person is not part of the construct; it is the baseline against which the construct becomes visible.

\section{A measurement model for AI-allowed items}
\label{sec:model}

\subsection{Cheap strategies and the AI floor}

\begin{definition}[Cheap strategy]
A \emph{cheap strategy} $s$ is a way of answering an item that requires effort no greater than a bound $\varepsilon$ and uses none of the target skill. The set of such strategies is $\mathcal{S}_\varepsilon$. Examples are sending the AI's reply to the pasted task unchanged (the \emph{copier}), pasting the task in pieces, relaying each new message to the AI, and making purely mechanical edits to the AI's reply.
\end{definition}

\begin{definition}[AI floor]
Let $m$ denote the pinned model and its instructions. For item $i$, let $\pi_i(s;m)$ be the probability that strategy $s$ produces a passing response. The item's \emph{AI floor} is the success rate of the strongest cheap strategy:
\begin{equation}
c_i(m) = \max_{s \in \mathcal{S}_\varepsilon} \pi_i(s;m).
\label{eq:floor}
\end{equation}
\end{definition}

\begin{assumption}[Knowledge or fallback]
A person who can do the human part of item $i$ passes it; a person who cannot falls back on a cheap strategy and passes with probability $c_i$. The probability of being able to do the human part follows a 2PL, $F_i(\theta)$.
\end{assumption}

\begin{proposition}[The AI floor is a lower asymptote]
Under the knowledge-or-fallback assumption, the probability of passing item $i$ is
\begin{equation}
P_i(\theta) = F_i(\theta) + [1-F_i(\theta)]\,c_i = c_i + (1-c_i)\,F_i(\theta),
\end{equation}
which is the three-parameter logistic model \eqref{eq:3pl} with lower asymptote equal to the AI floor.
\end{proposition}

The analogy with guessing is exact in form but different in substance. A multiple-choice guess succeeds at a rate fixed by the number of options; an AI floor is set by what the model can do with the item, and can be anywhere between 0 and 1. On an item the AI solves alone, $c_i \approx 1$ and every candidate passes.

\subsection{What the floor costs}

\begin{proposition}[Information retained]
\label{prop:info}
Let $I^0_i(\theta) = a_i^2 F_i(\theta)[1-F_i(\theta)]$ be the information of the item without a floor. With floor $c_i$, the item's information is
\begin{equation}
I_i(\theta) = \kappa_i(\theta)\, I^0_i(\theta), \qquad \kappa_i(\theta) = 1 - \frac{c_i}{P_i(\theta)}.
\label{eq:kappa}
\end{equation}
Hence $0 \le \kappa_i(\theta) \le 1-c_i$; at the item's difficulty ($\theta = b_i$), $\kappa_i = (1-c_i)/(1+c_i)$; and $\kappa_i(\theta) \to 0$ as $\theta \to -\infty$.
\end{proposition}

\begin{proof}
From $P_i = c_i + (1-c_i)F_i$, we have $P_i' = (1-c_i)\,a_i F_i(1-F_i)$ and $1-P_i = (1-c_i)(1-F_i)$. Substituting into \eqref{eq:info},
\[
I_i = \frac{(1-c_i)^2 a_i^2 F_i^2(1-F_i)^2}{P_i\,(1-c_i)(1-F_i)} = \frac{(1-c_i)F_i}{P_i}\, a_i^2 F_i (1-F_i) = \frac{P_i - c_i}{P_i}\, I_i^0 .
\]
Since $c_i \le P_i \le 1$, $\kappa_i \in [0, 1-c_i]$. At $\theta=b_i$, $F_i = 1/2$ and $P_i = (1+c_i)/2$, which gives $(1-c_i)/(1+c_i)$. As $\theta\to-\infty$, $F_i\to 0$ and $P_i \to c_i$.
\end{proof}

\begin{corollary}[Test length]
To keep the same measurement precision at the items' difficulty, a test whose items have floor $c$ needs $(1+c)/(1-c)$ times as many items as a test whose items have no floor.
\end{corollary}

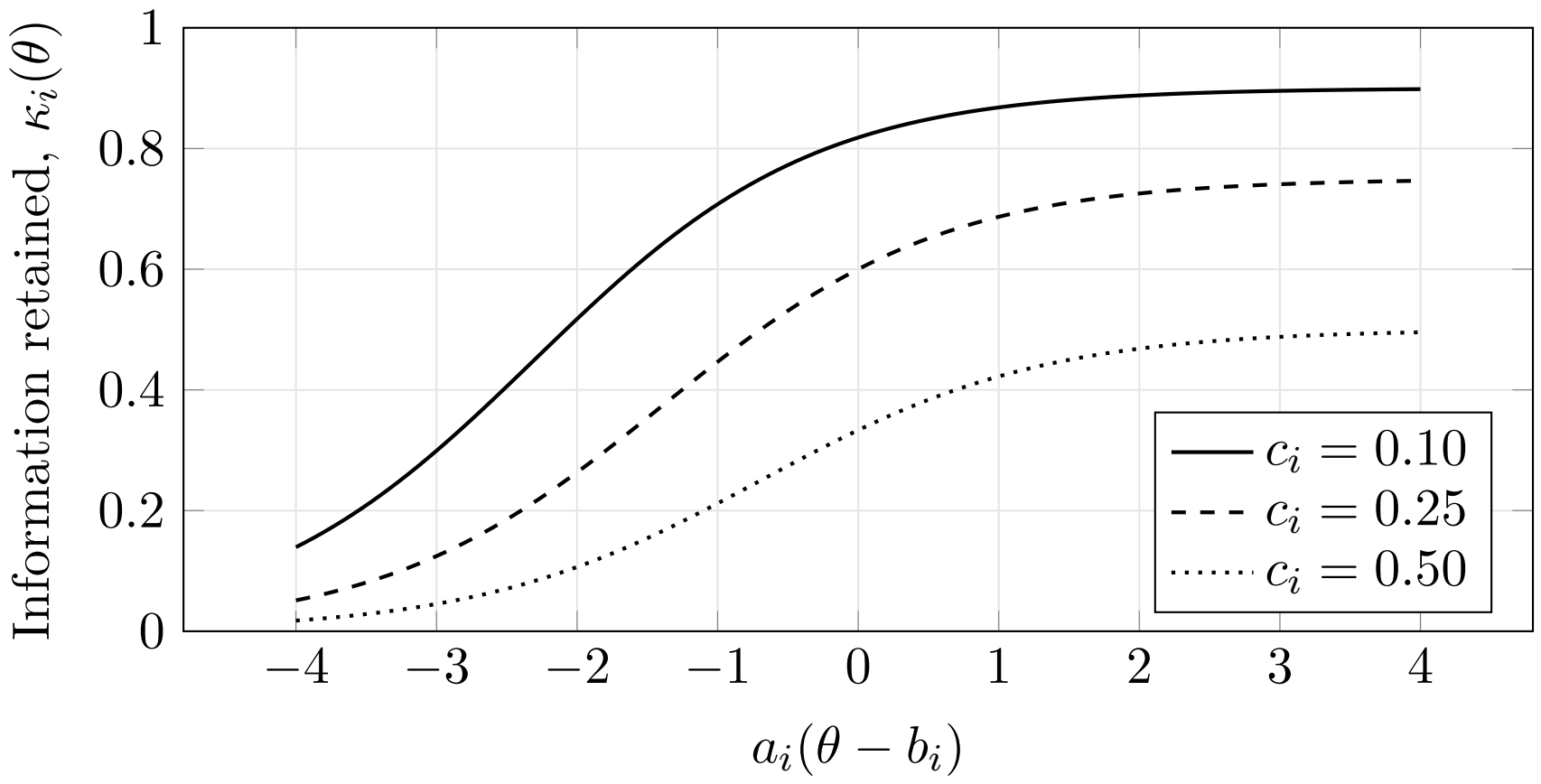
\begin{figure}[!htbp]
\centering
\begin{tikzpicture}
\begin{axis}[width=0.72\textwidth, height=6cm, domain=-4:4, samples=120,
  xlabel={$a_i(\theta-b_i)$}, ylabel={Information retained, $\kappa_i(\theta)$},
  ymin=0, ymax=1, legend pos=south east, legend cell align=left, grid=major, grid style={gray!20}]
\addplot[thick, black] {1 - 0.1/(0.1 + 0.9/(1+exp(-x)))};
\addplot[thick, black, dashed] {1 - 0.25/(0.25 + 0.75/(1+exp(-x)))};
\addplot[thick, black, dotted] {1 - 0.5/(0.5 + 0.5/(1+exp(-x)))};
\legend{$c_i=0.10$, $c_i=0.25$, $c_i=0.50$}
\end{axis}
\end{tikzpicture}
\caption{The share of an item's information about the person that survives an AI floor $c_i$ (Proposition~\ref{prop:info}). Information is lost most for people below the item's difficulty.}
\label{fig:kappa}
\end{figure}

Figure~\ref{fig:kappa} and Table~\ref{tab:floor} show the size of the effect. An item with a floor of one half keeps a third of its information at its own difficulty; three such items are needed to do the work of one floor-free item. Below the item's difficulty the loss is worse: the item cannot distinguish a person who lacks the skill from one who has a little of it, since both pass through the floor at the same rate.

\begin{table}[!htbp]
\centering\small
\caption{The cost of an AI floor at the item's difficulty.}
\label{tab:floor}
\begin{tabular}{@{}lcccccc@{}}
\toprule
AI floor $c_i$ & 0.05 & 0.10 & 0.20 & 0.25 & 0.30 & 0.50 \\
\midrule
Information retained, $(1-c_i)/(1+c_i)$ & 0.90 & 0.82 & 0.67 & 0.60 & 0.54 & 0.33 \\
Items needed for equal precision, $(1+c_i)/(1-c_i)$ & 1.11 & 1.22 & 1.50 & 1.67 & 1.86 & 3.00 \\
\bottomrule
\end{tabular}
\end{table}

A graded version follows the same logic. If an item is scored out of $M$ marks and the strongest cheap strategy earns $\mu_i$ in expectation, the \emph{headroom} $h_i = 1 - \mu_i/M$ bounds the difference any group of skilled candidates can show over it: the expected score of any candidate is at least $\mu_i$, since the cheap strategy is open to everyone.

\subsection{Estimating the floor before use}
\label{sec:estimation}

A multiple-choice guessing parameter is estimated from response data after an item is fielded. An AI floor can be estimated before any candidate sees the item, because the cheap strategies can be simulated with the pinned model itself. We call this simulation the \emph{Copy \& Paste Test} (\cp).

Let $n$ independent runs of strategy $s$ on item $i$ yield $k$ passing responses, so that $k \sim \mathrm{Binomial}(n, \pi_i(s))$. The exact upper confidence bound at level $1-\alpha$ \citep{clopper1934} is
\begin{equation}
U(k,n;\alpha) = B^{-1}(1-\alpha;\; k+1,\; n-k),
\label{eq:cp}
\end{equation}
where $B^{-1}$ is the quantile function of the beta distribution. When $k=0$ this reduces to $U = 1-\alpha^{1/n}$, approximately $3/n$ at $\alpha = 0.05$ \citep{hanley1983nothing}. For reporting, the Wilson interval \citep{wilson1927probable} is a good two-sided alternative.

\paragraph{Admission rule.} With a design ceiling $\tau$ on the floor, admit item $i$ only if
\begin{equation}
\max_{s \in \hat{\mathcal{S}}_i}\; U\!\big(k_s, n_s;\; \alpha/|\hat{\mathcal{S}}_i|\big) \;\le\; \tau ,
\label{eq:admit}
\end{equation}
where $\hat{\mathcal{S}}_i \subseteq \mathcal{S}_\varepsilon$ is the set of cheap strategies run on item $i$, and the division of $\alpha$ by their number, $|\hat{\mathcal{S}}_i|$, is a Bonferroni adjustment for testing several strategies. A second condition guards the other side: a simulated or real careful user must be able to reach full marks, so that the item's difficulty is not merely impossibility.

\paragraph{How many runs?} Even if no run of the cheap strategy passes, the floor can only be bounded as tightly as the number of runs allows. Setting $1-\alpha^{1/n} \le \tau$ gives the minimum number of runs:
\begin{equation}
n_{\min} = \left\lceil \frac{\ln \alpha}{\ln(1-\tau)} \right\rceil .
\label{eq:nmin}
\end{equation}
Table~\ref{tab:runs} gives values. Five runs with no copier success show only that the floor is below 0.45; ten runs, below 0.26; showing a floor below 0.10 takes 29 clean runs. Small simulation budgets, common in early design work, certify only loose ceilings.

\begin{table}[!htbp]
\centering\small
\caption{Upper 95\% bounds on the AI floor from $n$ runs with $k$ passes (exact), and runs needed to certify a ceiling $\tau$ when no run passes.}
\label{tab:runs}
\begin{tabular}{@{}lcccccc@{}}
\toprule
Runs $n$ & 5 & 10 & 10 & 20 & 20 & 50 \\
Passes $k$ & 0 & 0 & 1 & 0 & 1 & 0 \\
Upper bound $U$ & 0.45 & 0.26 & 0.39 & 0.14 & 0.22 & 0.06 \\
\midrule
Ceiling $\tau$ & 0.50 & 0.30 & 0.25 & 0.20 & 0.10 & 0.05 \\
Runs needed, $n_{\min}$ & 5 & 9 & 11 & 14 & 29 & 59 \\
\bottomrule
\end{tabular}
\end{table}

\subsection{The strongest cheap strategy, not the plainest}

Equation \eqref{eq:floor} takes a maximum over strategies. An item admitted against one strategy $s_0$ alone is admitted against a lower bound on its floor: $\pi_i(s_0) \le c_i$, and the shortfall $c_i - \pi_i(s_0)$ can be large. The design problem is adversarial. Its solution is to enumerate the cheap strategies a candidate could adopt within the effort bound, simulate the strongest, and treat any strategy not yet simulated as a threat to validity. The effort bound $\varepsilon$ matters: a strategy that requires reading the materials closely and catching what the AI got wrong is not cheap, because that is the skill.

\subsection{Three kinds of item}

Items differ in why their floor is low, and the reason predicts how long the floor will stay low.

\begin{proposition}[Structural zero]
Suppose a passing response on item $i$ requires either (a) information that first becomes available at time $t_0$ after the strategy has submitted its inputs to the AI, or (b) an action that the strategy does not perform (for example, choosing between tools or stopping an automated process). Then $\pi_i(s;m) = 0$ for every model $m$ and every strategy $s$ whose inputs are fixed before $t_0$ and which performs no such action.
\end{proposition}

\begin{proof}
Under (a), the strategy's response is a function of inputs that do not contain the information the passing response requires. Under (b), the passing condition is not met by any response the strategy produces. Neither argument depends on $m$.
\end{proof}

The proposition covers only strategies of a given form. A cheap strategy that relays each new message to the AI is not covered, so an item of type (a) is safe against it only if what the AI does with the new information still fails. We therefore distinguish:

\begin{description}[leftmargin=1.2em,style=nextline]
\item[Structural items.] The floor is held down by the structure of the task, as in the proposition. The floor does not depend on the model, at least for strategies of the covered form.
\item[Habit-based items.] The floor is held down because the model, left alone, makes a mistake a careful person would not send. The floor is $c_i(m) = 1 - \eta_i(m)$, where $\eta_i(m)$ is the rate of the habit; it depends on the model and, as Section~\ref{sec:surface} shows, on the item's surface. A cheap strategy that removes the habit's traces mechanically raises the floor towards one.
\item[Residue items.] The person's understanding is observed with the AI switched off (Section~\ref{sec:residue}). The floor does not depend on the model, because the model is absent, but on whether a cheap strategy lets the person answer without the skill.
\end{description}

\subsection{Model change and shelf life}

Because $c_i$ is indexed by $m$, a change of model or instructions changes every habit-based item's floor in an unknown direction. A model change is therefore a change of test form. It requires re-estimating every floor by the \cp, retiring items that no longer satisfy \eqref{eq:admit}, and linking the person scale across forms through anchor items \citep{kolen2014equating}. Items are retired because their floor has risen, not kept because they have become ``harder'' for the AI.

\subsection{Surfaces are not incidental for the AI}
\label{sec:surface}

Item-generation theory assumes that the item parameters depend on radicals only: $(a_i, b_i) = g(R_i)$. With an AI in the item, the floor depends on the radicals $R_i$, the incidentals $Z_i$ and the model $m$:
\begin{equation}
c_i = c(R_i, Z_i, m).
\end{equation}
Two isomorphs that share radicals can therefore have different floors, so that a change which is incidental for people can act as a radical for the AI. LLM benchmarks show the same sensitivity to template details \citep{mirzadeh2024gsmsymbolic}. Isomorphs are exchangeable for people only if each one's floor is estimated, which means the \cp{} must be repeated for every generated variant, not once per item model.

\subsection{The AI's randomness as a facet}

A pinned model is still stochastic: the same input can produce different outputs. Each person meets one draw of the AI's behaviour on each item. In the language of generalisability theory this adds a facet $a$, nested within person and item, to the design. With raters $r$ marking each response,
\begin{equation}
E\rho^2 = \frac{\sigma^2_p}{\sigma^2_p + \big(\sigma^2_{pi} + \sigma^2_{a:pi} + \sigma^2_{r:pia}/n_r\big)/n_i}.
\label{eq:gai}
\end{equation}
In operational data $\sigma^2_{a:pi}$ is confounded with the residual, since each person meets the AI once per item. It can be estimated separately from the \cp, which replays the same input many times. The number of items needed for a target coefficient $\rho^*$ is
\begin{equation}
n_i \;\ge\; \frac{\rho^*}{1-\rho^*}\cdot\frac{\sigma^2_{pi} + \sigma^2_{a:pi} + \sigma^2_{r:pia}/n_r}{\sigma^2_p}.
\end{equation}
Performance assessment already needs many tasks because $\sigma^2_{pi}$ is large \citep{shavelson1993sampling}; the AI's randomness adds to it.

\section{Foundations for scoring}
\label{sec:scoring}

\subsection{Marking the work as work}

Let $D$ be the deliverable a candidate sends and $L$ the record of how it was produced. We score the work through a function of the deliverable alone:
\begin{equation}
W = w_i(D), \qquad \text{so that } D = D' \;\Rightarrow\; w_i(D) = w_i(D') \text{ whatever } L \text{ and } L' \text{ are.}
\end{equation}
Four reasons support this \emph{authorship neutrality}. First, accountability: the person who sends work is responsible for it however it was produced, as employers, courts and publishers now hold. Second, the recipient sees only $D$; the employer's question is whether $D$ is right. Third, the joint product is the ``effect with'' the technology, which is the performance of interest (Section~\ref{sec:theory}). Fourth, the alternative, marking down work for being AI-drafted, rewards rewriting rather than judgement, introduces construct-irrelevant variance, and cannot be applied consistently, since authorship is not reliably detectable.

Authorship neutrality does not mean the person's contribution goes unmeasured; it means the contribution is measured elsewhere. It shows in three places: in the work itself, wherever the AI alone would have produced something that fails, so that a passing $D$ is evidence of the person's contribution in proportion to the item's headroom; in the residue; and in the record of how the AI was used. Because the construct includes the person's actions in working with the AI (Section~\ref{sec:theory}), some evidence is read from $L$: what the candidate gave the AI, what they told it about the task, and which tool received which data. This evidence concerns observable acts of use, never a judgement of who drafted a passage, so neutrality about authorship is kept. A sitting in which the AI is not used yields no evidence about the construct, and the item's competency receives a failing result.

\subsection{The residue: understanding one's own work}
\label{sec:residue}

After the work is submitted and locked, the AI is switched off and the candidate answers short questions about their own deliverable. The answer is judged right if it correctly describes what the candidate's own work says or does, even when that work is wrong; the fault in the work is marked in the work. This separates two errors that are otherwise confounded: sending wrong work, and not knowing what one sent. The residue score $R = r_i(\text{answers})$ is a local version of Salomon's ``effect of'': not lasting learning, but the understanding a person must have to stand behind a specific piece of work. It is evidence for a learning objective of its own (Section~\ref{sec:bands}), placed with the skill the item tests rather than with its minimum, so that work a person cannot explain does not reach the passing standard. The nearest precedent is asking programming students questions about code they submitted \citep{lehtinen2021qlc}; about a third of students struggled to explain their own code, including many whose code passed every test \citep{lehtinen2021explain}.

\subsection{Learning objectives, bands and serious failures}
\label{sec:bands}

Each item states what it tests as a small set of learning objectives, and its Answer Key states, for each objective, the evidence that shows it, as points marked yes or no. An objective is met only when all of its evidence is present; there is no partial credit within an objective. The objectives are grouped into ordered bands: a basic band, the minimum for this kind of task; a skilled band, the skill the item tests; and, where the item has them, an upper band of harder extras. A band counts only when every objective in it, and every band below it, is met. If $\mathcal{O}_{ik}$ is the set of objectives in band $k$ of item $i$ and $y_{po} \in \{0,1\}$ records whether person $p$ met objective $o$, the band reached is
\begin{equation}
\ell_{pi} = \max\Big\{k : y_{po} = 1 \text{ for every } o \in \mathcal{O}_{i1} \cup \dots \cup \mathcal{O}_{ik}\Big\},
\end{equation}
with $\ell_{pi} = 0$ when the basic band is not met. This is a conjunctive rule of the kind \citet{haladyna1999conjunctive} compare with compensatory ones, applied to the objectives of one item rather than to the parts of a test. Which objectives sit in which band is fixed for each type of item, so that items of the same type are graded alike, and the passing standard is the skilled band: reaching it is what Section~\ref{sec:model} calls a passing response.

What counts as a serious failure needs a principled test, or the list grows arbitrarily. We propose four conditions. An error $e$ is serious if and only if (i) it is in the delivered work or in the recorded use of the AI, such as restricted data sent to an unapproved tool; (ii) the recipient would act wrongly on it, or the organisation would be harmed or embarrassed by it; (iii) it can be pointed to; and (iv)
\begin{equation}
\Pr(e \mid \text{competent person, working without AI}) \approx 0 .
\end{equation}
Condition (iv) anchors the standard in ordinary professional work rather than in the AI: the question is not whether the AI tends to make the error, but whether a competent professional would ever send it. A serious failure is not compensated by correct work elsewhere in the item, as with critical errors in licensure examinations \citep{buckendahl2012dce}; fully compensatory pass rules have also been questioned for clinical-skills examinations \citep{clauser1996compensatory}. Rather than subtracting marks, a serious failure sets a ceiling on the item's result: an error that would mislead the recipient or harm the organisation, a \emph{critical} failure, holds the result below the basic band, and a lapse of care, such as an unfilled template field, holds it below the passing standard. One fault counts once: a fault that already leaves its own objective unmet sets no further ceiling. A leak of private information is the exception: it always holds the result below the basic band, so that it always costs more than a careless leftover. Where the marker cannot tell with confidence whether an error was made or an objective met, it refers the point to the test owner, and the point is not held against the candidate until it is ruled on. Each item measures one competency and counts only for it, which keeps each item's evidence unidimensional and stops one fault from failing several claims. The candidate is never stopped; a failing result is a result, not a removal.

Where a number is reported, the item's score rises with the band reached and, within a band, with the share of the item's objectives met, and is then capped by the most serious failure found:
\begin{equation}
S_{pi} = \min\big(q(\ell_{pi}, \bar{y}_{pi}),\; u(F_{pi})\big),
\label{eq:itemscore}
\end{equation}
where $\bar{y}_{pi}$ is the share of the item's objectives met, $q$ increases in both arguments, and $u(F_{pi})$ is the ceiling set by $F_{pi}$, the most serious failure recorded, with no ceiling when there is none. No marks are deducted, so correct work is never wiped out by penalties. Time is recorded but never enters $S$: it is reported beside the result, ranked among the candidates being compared, such as those in the same recruitment round, and only where the passing standard is met. A candidate who sends the AI's output unread is always the fastest, so scoring speed would reward copying and penalise checking.

\subsection{Checking as signal detection}

Reliability, catching what is wrong in AI-assisted work, is a detection task and is naturally modelled by signal detection theory \citep{green1966sdt,macmillan2005detection}. Suppose a checking item presents $n_E$ unsupported statements and $n_T$ true ones, and the candidate corrects some. With hit rate $H$ (errors corrected) and false-alarm rate $\mathit{FA}$ (true statements changed), sensitivity and criterion are
\begin{equation}
d' = \Phi^{-1}(H) - \Phi^{-1}(\mathit{FA}), \qquad c_{\mathrm{SDT}} = -\tfrac12\big[\Phi^{-1}(H) + \Phi^{-1}(\mathit{FA})\big].
\end{equation}
Because a checking item holds few errors, observed rates of 0 or 1 are common and make $\Phi^{-1}$ infinite; the log-linear correction adds one half to each count, for example $H = (\text{hits} + 0.5)/(n_E + 1)$ \citep{hautus1995corrections}, and sensitivity is best estimated across several items. The criterion that maximises expected value sets the likelihood-ratio threshold at
\begin{equation}
\beta^* = \frac{\Pr(\text{true})}{\Pr(\text{error})}\cdot\frac{V_{\mathrm{CR}} + C_{\mathrm{FA}}}{V_{\mathrm{H}} + C_{\mathrm{M}}},
\end{equation}
where $V$ are the values of correct rejections and hits and $C$ the costs of false alarms and misses. This formalises calibrated trust \citep{lee2004trust}: over-reliance is a criterion too lenient for the stakes, and reflexive suspicion a criterion too strict. A test that scores only hits rewards suspicion; scoring false alarms as well, by placing true and surprising statements beside the errors, rewards calibration. In an item's Answer Key the two sides are separate objectives, catching the errors and leaving correct content alone, and both are needed for the skilled band. The same logic covers acting on rules: holding, refusing or cancelling where no written rule is broken is a false alarm, and the right response is to raise the matter with the person whose decision it is.

\subsection{Decision rights}

A related condition applies to work outside a person's field. Let $\Delta_i$ be the decisions an item's situation contains, each with an owner. The deliverable meets the decision-rights condition if, for every decision $\delta \in \Delta_i$ whose owner is not the candidate's role, $D$ does not make $\delta$ and passes it to its owner with what is known. Language models readily recommend; a person working with one must know when a recommendation is not theirs to send.

\subsection{Answer Keys as precedent}

Answer Keys for constructed responses are usually written item by item. We instead derive each Answer Key, the evidence for each of the item's objectives $\mathcal{O}_i$, from a set of named marking principles $\Pi$:
\begin{equation}
K_i = K(\Pi, \mathcal{O}_i, \text{materials}_i).
\end{equation}
When a mark is in doubt, because the automated marker records that it is unsure or a reviewer disagrees with it, the doubt is settled by an existing principle; if none covers it, the test owner rules, and the ruling is added to $\Pi$, so that later markers apply it rather than guess. $\Pi$ grows monotonically, every Answer Key can be traced to principles, and once a ruling is made, like cases are marked alike across items. The approach resembles the development of codebooks in team qualitative analysis \citep{macqueen1998codebook}, the amendment of mark schemes at examiners' standardisation meetings, automated rubric refinement from human--LLM disagreements \citep{chu2025gradehitl}, and case-based reasoning for AI policy \citep{feng2023case}. It also addresses the finding that marking criteria drift as people judge outputs \citep{shankar2024validators}: the drift is captured as explicit principles rather than left implicit.

The automated marker is one rater among others, and its severity $\lambda_j$ in the many-facet model \eqref{eq:mfrm} can be estimated against human raters and monitored across model versions. One bias deserves particular attention: language models favour text they generated themselves \citep{panickssery2024self,ye2025justice}; \citet{zheng2023judging} saw early signs of this but could not confirm it with their data. When much of the marked work was drafted by an AI, the marker should come from a different model family.

\subsection{From items to levels}

Competencies fall into two groups: \emph{value} (does AI make the person's work better?) and \emph{trust} (can their AI-assisted work be relied on?). Each competency is reported as an area whose result combines the results of its items, and a test is passed only when every area in it is passed:
\begin{equation}
\text{pass} \iff \min_a S_a \ge \gamma,
\end{equation}
where $S_a$ is the result in area $a$ and $\gamma$ the passing standard. In the areas concerned with risk, a critical failure in any item holds the area below $\gamma$, so that a good average cannot hide a single leak of data or an unsafe step. Trust is a floor: AI that makes work faster but unreliable does not make a person proficient. Results may also be reported against other candidates, but only within the same area and level, and only once the comparison group is large enough to be stable.

\section{The framework in practice: the SIAA Standard}
\label{sec:siaa}

The SIAA Standard is a work-sample test of applied AI skill researched and developed by the Straits Institute for Applied AI. Candidates do real tasks from many professions, general and specialist, with an AI assistant on the test screen. Six competencies are reported, three for value (Acceleration: faster and better at known work; Expansion: competent outside one's field while knowing one's limits; Delegation: keeping control of AI that acts on its own) and three for trust (Safety: keeping data and tool rules; Reliability: outputs that can be relied on; Accountability: work colleagues can rely on). Levels run from Novice through Developing to three certified levels, Proficient, Advanced and Expert. This section describes the design as it stood in November 2026; it is still changing.

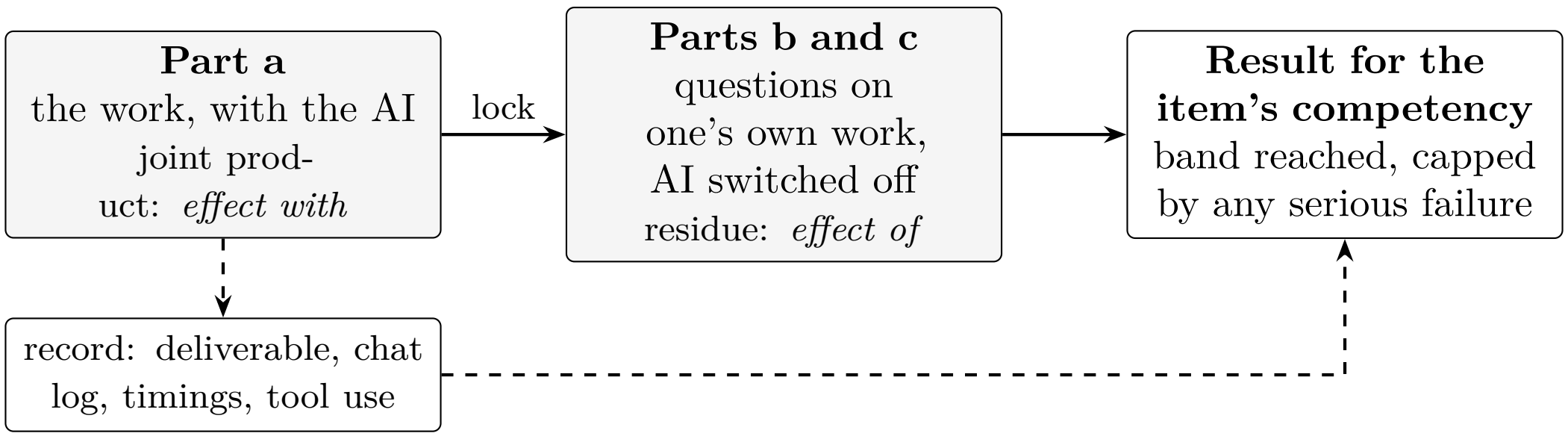
\begin{figure}[!htbp]
\centering
\begin{tikzpicture}[font=\small, box/.style={draw, rounded corners=2pt, align=center, minimum height=1.0cm, text width=3.6cm}, arr/.style={-{Stealth[length=2mm]}, thick}]
\node[box, fill=gray!8] (a) {\textbf{Part a}\\ the work, with the AI\\ \footnotesize joint product: \emph{effect with}};
\node[box, fill=gray!8, right=1.1cm of a] (bc) {\textbf{Parts b and c}\\ questions on one's own work,\\ AI switched off\\ \footnotesize residue: \emph{effect of}};
\node[box, right=1.1cm of bc] (y) {\textbf{Result for the item's competency}\\ band reached, capped\\ by any serious failure};
\draw[arr] (a) -- node[above]{\footnotesize lock} (bc);
\draw[arr] (bc) -- (y);
\node[box, below=0.7cm of a, text width=3.6cm, fill=white] (log) {\footnotesize record: deliverable, chat log, timings, tool use};
\draw[arr, dashed] (a) -- (log);
\draw[arr, dashed] (log.east) -| (y.south);
\end{tikzpicture}
\caption{The structure of a question with follow-ups.}
\label{fig:screen}
\end{figure}

Questions come in several formats, and each item carries its own defence against copying, because a floor held down only by the AI's habits is unstable (Section~\ref{sec:habit}): follow-up questions with the AI switched off, information that arrives during the task, requests or an agent's actions that appear one at a time, or errors planted in the AI's draft. In some formats the AI is closed throughout and only the candidate's prompt is marked. Figure~\ref{fig:screen} shows a common format: in part a, the candidate does one piece of work with the AI; in parts b and c, the AI is switched off and the candidate answers questions about their own work. Everything is recorded: the deliverable, the full chat with timings, messages that arrive during the task, and, where there are several AI tools, what was sent to which.

\paragraph{A pinned, unscripted AI.} Every candidate faces the same model with the same instructions, but the model is not scripted. It behaves as it behaves, including its mistakes, so that the test measures how people work with the AI that exists rather than with one made to fail. The single exception is the Reliability format in which the candidate checks the AI's draft, where the test of checking needs something to check: there, errors are inserted into the AI's draft by code after the model replies. Two findings shaped this rule. Instructed to state a falsehood about the candidate's materials, the model refused, and sometimes explained why, revealing the plant; and an error planted in every item would make every score partly a Reliability score. Planted errors follow error types that published studies find common in machine-written summaries: unsupported addition, meaning content the source does not contain \citep{maynez2020faithfulness,pagnoni2021frank,williams2025edsummaries}, and omission of relevant content \citep{williams2025edsummaries}; invented numbers, by contrast, are rare in long financial summaries \citep{cao2024financial}.

\paragraph{Item life cycle.} Each item is built afresh through a fixed sequence of steps: a question type, set by level, competency and format, which carries the item's learning objectives; a real task from a profession; the item, and then separately its Answer Key, with every piece of evidence traced to a line of the materials; checks of fairness and accuracy; and a trial sitting by a person before the marking and grading are fixed. The materials must contain every fact the work needs and nothing that points at what is marked, because the AI reads the materials too: a sentence that signals what the candidate should notice is noticed by the AI as well. Before use, each item is screened by the \cp{}: five runs of the pinned model on the task alone are marked and graded like any response, and unless the copier fails or scores clearly lower than careful work, the item is revised. Five clean runs bound the floor only below 0.45 (Table~\ref{tab:runs}), so the screen catches items the AI can do alone but does not certify a low floor; certification to a design ceiling (Section~\ref{sec:estimation}) is still to be done. The screen is repeated when the model changes, and an item refilled for another profession would need its own screen (Section~\ref{sec:surface}).

\paragraph{Scoring.} Scoring follows Section~\ref{sec:scoring} throughout: every item has its learning objectives and bands, and every change to the marking principles is logged.

\section{Evidence from simulation}
\label{sec:evidence}

\subsection{Method}

Between 24 and 30 October 2026, during the design of the first 16 bank questions and a 10-task starting-level check, we ran the \cp{} with the test's pinned model, a current commercial language model. About 700 cheap-strategy runs were saved: 326 full pastes, 260 pastes in pieces, 50 single pastes and 60 relays on live tasks, and 10 reworded pastes, with 17 simulated careful-user runs. Runs reached the model at default sampling settings, by a slightly different route from the candidate page's, and we did not test whether the route matters. Every reply was read and marked against the Answer Key by the design team's AI assistant, with mechanical traces also detected by pattern matching. Items were revised against the same model until they passed, so rates for final versions are optimistic, and most conditions had 5 or 10 runs, which by Table~\ref{tab:runs} certifies only loose bounds.

\subsection{What the AI neutralised}

Table~\ref{tab:neutral} lists design features that did not lower the floor. Each is a familiar way to make a task harder, and each was handled by the model alone.

\begin{table}[!htbp]
\centering\small
\caption{Design features the pinned model neutralised in the \cp.}
\label{tab:neutral}
\begin{tabularx}{\textwidth}{@{}P{4.5cm}LP{2.6cm}@{}}
\toprule
Feature & What the model did & Runs \\
\midrule
Facts placed in a side panel or behind a button & Used them, once they were pasted & full marks, 5 of 5 on three tasks \\
A rule written in the materials & Kept it & 10 of 10 \\
A multi-step calculation & Got it right & 5 of 5 \\
Two figures in the materials that conflict & Noticed the conflict & 10 of 10 \\
Errors in a draft, with the source also given & Corrected all of them & 10 of 10; 5 of 5 \\
Pressure or disagreement in the materials & Took the careful course & every run, two tasks \\
Sentences pointing at what the candidate should notice & Repeated the pointer and acted on it & see text \\
An AI agent given weak instructions & Acted safely anyway & every run, three instruction sets \\
An instruction to the model to state a falsehood & Refused & 0 of 5 complied \\
\bottomrule
\end{tabularx}
\end{table}

Two results deserve comment. Removing two pointer sentences from one task's materials changed the model from mostly handing a decision back to its owner to recommending a course of action in 7 or 8 of 10 runs: the sentences had done the candidate's thinking for the AI. And an agent run overnight on a scripted stream of customer and system messages, including refund demands and fake instructions to cancel orders, made no refunds, no wrong cancellations and followed no fake instruction, whether its instructions were careful, written by the AI, or a single line. Weak instructions did no harm because the agent did not need them; a test of delegation cannot rest on the consequences of weak instructions while the model is this cautious.

\subsection{Habit-based floors are unstable}
\label{sec:habit}

Habit-based items depend on the rate at which the model leaves a fault a careful person would remove. Across tasks and versions, those rates ranged from 0 to 10 in 10. On one item, adding six facts that closed gaps in the materials, a change with no bearing on the human task, cut the model's habit rate from 9 in 10 to 4 in 10 in full pastes, while pasting in pieces still produced it 10 in 10. This is the surface sensitivity of Section~\ref{sec:surface}, observed.

\subsection{Plain copier against strongest cheap strategy}

Ten of the 16 bank items met the admission rule against the plain copier through the work itself, two of them structural items admitted by construction without runs; five rested on the residue alone, and one had to be run without the AI because no version with the AI could be made to defeat the copier. Re-evaluated against the strongest cheap strategies we could construct from the same runs, only three of the ten remained admissible in every condition (the two structural items and one checking item), and three more only against a candidate who pastes once rather than relaying each message. The other four fell back on the residue. Admission against the plain copier overstated the margin of most habit-based items, as the maximum in \eqref{eq:floor} predicts.

\subsection{Grading against copiers}
\label{sec:gradingtest}

The rules of Section~\ref{sec:bands} were tested in November 2026 in four stages, while they were being decided. First, 155 plain-copier runs on 12 items of the current bank, already marked, were graded under the rules. One reached the passing standard, on an item version since revised; of the rest, 85 reached the basic band, correct and safe but copied, and 69 fell below it. Second, five kinds of simulated candidate (careful, careless at the finish, over-sharing, over-checking, and not using the AI) sat six items: 30 sittings, each marked by the operational automated marker and then graded. Careful work passed five of the six items; sittings without AI use received a failing result on all six. The simulated poor candidates were less informative than planned: told to over-share or over-check, the AI playing them still mostly did the careful thing, so poor behaviour is better tested by planting a fault in otherwise good work. Third and fourth, whole tests were graded, sat by simulated candidates, most of them strong work with one fault put in by hand, and marked by the operational marker. With one item per area, a single careless fault in one item failed its area, and with it the test; with two shorter items per area, a single careless fault was averaged out and the area passed, while a leak of private information still failed its risk area. The tests also changed the rules: one fault now counts only once, an unsure mark waits for a ruling, and a leak, which at first scored no lower than a forgotten placeholder, now always holds its item below the basic band.

\section{Discussion}
\label{sec:discussion}

\subsection{What the framework adds}

Most of the components have precedents: the lower asymptote of the 3PL, adversarial filtering, item models, standardised counterparts, critical errors, signal detection, and evolving codebooks. The framework's contribution is to connect them for a test in which AI is the tool and the use of AI is the skill. It identifies the AI floor as the parameter that threatens such tests; shows what the floor costs in information and in test length; shows that, unlike a guessing parameter, the floor can be estimated before an item is used, and how many runs that takes; shows why it must be estimated against the strongest cheap strategy and re-estimated after any change of model or surface; adds the AI's randomness to the error model; and grounds scoring in the distinction between the joint product and the person's residue, with grading rules tested against copiers.

\subsection{The validity argument}

Table~\ref{tab:validity} sets out the inferences from a candidate's performance to a decision \citep{kane2013validating}, what each assumes, and the evidence so far. Only the scoring inference, and the placing of the passing standard, have support so far, and that support comes from simulation.

\begin{table}[!htbp]
\centering\small
\caption{The validity argument: what must hold, and what is known.}
\label{tab:validity}
\begin{tabularx}{\textwidth}{@{}P{2.3cm}LLL@{}}
\toprule
Inference & Assumption & Evidence so far & Needed \\
\midrule
Scoring & Scores reflect the person's contribution, not the AI's & Floors estimated by the \cp; items screened with five runs; marker checks on a few records; grading tests on 155 copier runs, 30 simulated sittings and whole tests (Section~\ref{sec:gradingtest}) & Floors certified with enough runs; real copiers and careful users; human raters against the automated marker \\
Generalisation & Results hold across items, variants and AI draws & Grading tests showing one item per area to be too few (Section~\ref{sec:gradingtest}) & A generalisability study including the AI facet \eqref{eq:gai} \\
Extrapolation & Results reflect AI use in the real job & Tasks drawn from descriptions of real work & Comparison with supervisors' judgements of real work \\
Decision & Levels support fair certification decisions & Passing standard checked against simulated copiers (Section~\ref{sec:gradingtest}) & Standard setting \citep{cizek2007standard}; differential item functioning \citep{holland1988dif} \\
\bottomrule
\end{tabularx}
\end{table}

\subsection{Limitations}

All evidence comes from one model, simulated strategies and revised items. The knowledge-or-fallback assumption is a simplification: real candidates mix strategies, edit partially, and sometimes fail tasks they could do, just as examinees on matching tests often use strategies that break the random-guessing assumption of the knowledge-or-random-guessing model \citep{vanderven1992krg}. The residue may favour fluent readers and writers, which needs a fairness study. Floors certified with small simulation budgets are loose, and the five-run screen now applied to new items bounds a floor only below 0.45 (Section~\ref{sec:siaa}). The grading rules were tested on simulated candidates only, and the passing standard was placed by the design team and checked against copiers, not set by a standard-setting panel. And every result is tied to a model version that will be replaced.

\subsection{Research agenda}

Six studies follow directly. (1) A field study with instructed copiers, instructed light editors and careful users, to test whether simulated floors predict real ones. (2) A generalisability study with items, variants, AI draws and raters as facets. (3) A rater study placing the automated marker and trained humans on one many-facet scale. (4) A re-estimation of every floor under a newer model, to measure how fast habit-based floors move. (5) A study of how people actually use AI at work, to establish which cheap strategies are common and therefore which must be simulated. (6) An experiment in which managers judge the same work labelled as written by an employee or drafted by AI, to test whether authorship neutrality matches how recipients judge work, and whether they share the serious-failure conditions of Section~\ref{sec:scoring}. A survey of working adults covering (5) and (6) has been designed.

\section{Conclusion}

When a test allows AI, the first question to ask of every item is how well the AI does alone. We have shown that the answer enters the measurement model as a lower asymptote, the AI floor; that the floor costs information in a precise way; and that, because the AI can be simulated, the floor can be measured and controlled before any candidate sees the item. The simulation evidence also carries a warning: with a current model, many familiar ways of making a task hard are done by the AI alone. What remains to measure is what only the person brings: acting, noticing what arrives later, knowing whose decision it is, finishing work for the person who receives it, and understanding what they sent.

\section*{Declarations}

\paragraph{Use of AI tools.} An AI assistant (Claude, Anthropic) worked under the Institute's direction: it helped design and run the simulations, read and marked the runs, searched the literature, and drafted this manuscript. The authors take full responsibility for the content.

\paragraph{Competing interests.} The authors work for the Straits Institute for Applied AI, which is developing the SIAA Standard as a commercial assessment.

\paragraph{Data availability.} The simulation transcripts contain operational test items and are not released, in keeping with test-security practice \citep{aera2014standards,wollack2013security}. Aggregate results are reported in full above.

\FloatBarrier
\setlength{\bibsep}{3pt plus 1pt}
\bibliographystyle{plainnat}
\bibliography{references}

\end{document}